# Electromagnetic Grazing Anomaly: Energy Flux Resonance Behaviour and Brewster Angle Analogy

T. Rokhmanova[1,2,3], and A. V. Kats[1]

[1] O. Ya. Usikov Institute for Radiophysics and Electronics of the National Academy of Sciences of Ukraine, Ak. Proskury 12, 61085 Kharkiv, Ukraine,

[2] V.N. Karazin Kharkiv National University, Kharkiv 61077, Ukraine,

[3] Departamento de Física–CIOyN, Universidad de Murcia, Murcia 30.071, Spain.

E-mail: avkats@ire.kharkov.ua

**Abstract**

The diffraction of electromagnetic waves at the surface periodic structures accompanied by strong anomalous effects in different diffraction orders is considered in great detail for high-contrast interfaces. We restrict our discussion to the TM polarization of the incident wave (the magnetic field is orthogonal to the plane of incidence) and the simplest geometry when the plane of incidence is orthogonal to the grating grooves. The most attention is focused on the strong maxima and minima of the energy flux density accompanying specific grazing propagation of some diffraction order. Relation to other anomalies, both Rayleigh and the resonance ones is discussed as well.

**1. Introduction.**

It has been well known since the early 1900s that the light diffraction on metal gratings is accompanied by a number of strong spectral and angular anomalies which manifest themselves in the fast dependence of the intensities on the wavelength and/or angle of incidence. The pioneer work on the subject was performed by R. Wood in 1902 [1] with metal gratings. The first physical interpretation of some of the observed peculiarities was presented by Lord Rayleigh in [2]. He associated them with the branch points related to diffracted waves (i.e., with the transition from the outgoing wave to the evanescent (decaying) one and vice versa in different diffracted orders). Such an explanation is incomplete due to the fact that some Wood anomalies are to be attributed to the resonance excitation of the surface electromagnetic wave at the metal/air interface. Such interpretation was first proposed by U. Fano, [3]. The resonantly excited waves are called the surface plasmon polaritons (SPPs), [4]. The resonance anomaly is still widely discussed due to its perspective role in nanophotonics. Later, Wood caught site of one more anomaly relating to the anomalously high intensity of the grazing outgoing wave: "…the spectrum leaving at grazing emergence, which is the one which governs the appearance of the anomalous bands, is very bright."[5] Below the anomalies attributed to the grazing propagating waves are referred to as GA (Grazing Anomaly).

It is essential that the Rayleigh anomaly exists for an arbitrary interface and light polarization. However, it is much more pronounced for the high-dielectric contrast interface and for TM (transverse magnetic) polarization if the media we are dealing with are nonmagnetic. In what follows we restrict the consideration to the nonmagnetic case only. The results for the magnetic case can be obtained by replacing the dielectric permittivity, $\varepsilon$, with the magnetic permeability, $\mu$, and the TM polarization by the TE one and vice versa. The resonance anomaly can exist only for such interfaces that support surface electromagnetic waves (SEW). GA anomaly is rather universal and is well expressed for high contrast interfaces for TM polarization. To the best of our knowledge, it was first discussed theoretically in [6].

Consider briefly the main properties of these anomalies. The branch (Rayleigh) point anomaly is of general type, its position can be easily obtained from the Bragg diffraction conditions and it exists for arbitrary polarization and interfaces. However, it is more significant for metals under TM polarization. At the Rayleigh point the derivative of the diffracted wave intensity with respect to the wavelength or angle of incidence turns infinity. The resonance anomaly is less general because it is caused by existence of well-defined eigenmodes of the interface.[1] For isotropic and nonmagnetic dissipation-free media such surface-localized electromagnetic waves do exist under the conditions $\varepsilon < 0$, $\varepsilon_d > 0$, $\varepsilon_d + \varepsilon < 0$, where $\varepsilon$ and $\varepsilon_d$ denote dielectric permittivity of the metal and the adjacent dielectric, respectively. The SPP in-plane wavenumber, $Q$, $Q = Q(\omega) = \frac{\omega}{c}\sqrt{\varepsilon\varepsilon_d/(\varepsilon+\varepsilon_d)} > 0$, where $\omega$ is the (angular) frequency of the incident wave, exceeds the wavenumber of the adjacent dielectric volume wave with the same frequency, $k = k(\omega) = \omega\sqrt{\varepsilon_d}/c$, $Q > k$. The square root symbol stays for the main branch, so that $\sqrt{Z} = \sqrt{|Z|}\exp(i\phi/2)$ for $Z = |Z|\exp(i\phi)$ with $\phi \in [0, 2\pi)$. The SPP is TM polarized, i.e., if it propagates along the interface $z = 0$ in $Ox$ direction then its magnetic field, $\mathbf{H}$, is directed along $Oy$ direction, $\mathbf{H} = (0, H, 0)$, and the electric field, $\mathbf{E}$, lies in the $xOz$ plane, $\mathbf{E} = (E_x, 0, E_z)$. The space dependence of the SPP fields in the dielectric halfspace, $z \leq 0$, is given by the ansatz $\exp[iQx - ip(Q)z]$, where the function $p(\mathbf{q})$ is defined for arbitrary two-dimensional vector $\mathbf{q} = (q_x, q_y)$ so that,

$$p(\mathbf{q}) = \sqrt{k^2 - \mathbf{q}^2}, \qquad k = \sqrt{\varepsilon_d}\omega/c, \quad \text{Re, Im}\, p(\mathbf{q}) \geq 0. \quad (1)$$

In the specific case of SPP, the quantity $p(Q)$ is $z$-component of the wavevector in the dielectric and for dissipation-free media it is pure imaginary under the condition $\varepsilon + \varepsilon_d < 0$, $p(Q) = i|p(Q)|$, so that the field amplitude decays exponentially with increasing distance from the interface $z = 0$.

Recall, if the plane monochromatic electromagnetic wave with space dependence,

$$\mathbf{E}, \mathbf{H} \propto \exp[i(\mathbf{q}\cdot\mathbf{r}) + ip(\mathbf{q})z], \qquad \mathbf{q} = (q_x, q_y), \quad (2)$$

(here and further the time dependence is supposed to be of the form $\exp(-i\omega t)$ and is omitted) is incident on the interface from the dielectric medium located at negative $z$ values, $-\infty < z < \zeta(x)$, where the surface profile, $z = \zeta(x)$, presents periodic function with period $d$, $\zeta(x+d) = \zeta(x)$, then the electromagnetic field within the dielectric medium is the sum of spatial harmonics of the form,

$$\mathbf{E}_n, \mathbf{H}_n \propto \exp[i(\mathbf{q}_n\cdot\mathbf{r}) - ip(\mathbf{q}_n)z],$$
$$\mathbf{q}_n = \mathbf{q} + n\mathbf{g}, \quad \mathbf{g} = \mathbf{e}_x 2\pi/d, \quad n = 0, \pm 1, \pm 2, \ldots \quad (3)$$

where $\mathbf{e}_x$ is the unit vector directed along the $Ox$ axis. In other words, the diffracted field is given by the Floquet-Fourier expansion, [7, 9]. In (3) the sign minus before $p(\mathbf{q}_n)$ stays to satisfy the radiation boundary conditions at $z = -\infty$. Restriction of the outgoing waves (and evanescent decaying ones) within the whole halfspace $z \leq \zeta(x)$ corresponds to use of the Rayleigh hypothesis, [2], and is not restrictive even for rather deep gratings, see recent discussion in [7, 9, 10].

Consequently, if for some specific integer $n$ the condition $|\mathbf{q}_n| \simeq Q$ holds true, then for the appropriate polarization of this diffracted wave the resonance excitation of SPP takes place. It is significant that SPP is an evanescent wave and thus the magnitude of the corresponding diffracted order can exceed essentially that of the incident wave. Specifically, in the simplest geometry, when $\mathbf{q}$ is orthogonal to the grating grooves, $\mathbf{q} = (q, 0)$, $q > 0$, only TM component of the incident wave can excite the SPP.[2] Also, it worth mentioning that for the modulated interface the SPP resonance center experiences shift in comparison with the "naked" condition, $|\mathbf{q}_n| = Q$. However, the SPP resonance in the majority of experimental situations in visible and near-infrared spectral regions seems to be rather evident to attribute.

We would like to underline that the Rayleigh and the resonance anomalies are related to the specific and rather sharp dependence of the field *amplitudes* on the wavelength and angle of incidence. They can be considered on the basis of simple qualitative treatment. The treatment of the third mentioned Wood anomaly cannot be accomplished without a thorough theoretical investigation. This obstacle is caused by the fact that the field amplitude changes monotonically within the anomaly. It can be shown that the corresponding quasiresonance behavior is characteristic for the intensity, not for the field amplitude. The method for considering this and other diffraction anomalies analytically was presented in [9], see also a more detailed consideration in [11-13].

---

[1] We restrict our consideration to the interface of two homogeneous isotropic nonmagnetic media, e.g., metal and dielectric. If between these two media exists some third one (even very thin layer), then additional to SPP resonances can occur, [7]. For anysotropic media the resonance can be caused by other than SPP surface modes, e.g., Dyakonov ones, see [8] and citations therein.

[2] Noteworthy, in the simplest geometry the diffraction of TE and TM components of the incident wave are independent processes and thus can be considered separately.

## 2. Grazing incidence anomaly.

In this section, we present the brief summary of the results for the case of the simplest geometry which are essential for the further consideration. For the TM polarization of interest the magnetic field is orthogonal to the plane of incidence and thus possesses the $y$-component only, so that for the incident wave, $\mathbf{H}^i$, and for the Fourier-Floquet expansion of the diffracted field, $\mathbf{H}^D$, we have,

$$\mathbf{H}^i = \mathbf{e}_y H \exp[iqx + ip(q)z],$$
$$\mathbf{H}^D = \mathbf{e}_y \sum_{n=-\infty}^{\infty} H_n \exp[iq_n x - ip(q_n)z], \ z \le \zeta(x) \quad (4)$$

where $q_n = q + ng$. Note, the diffracted field in (4) and below in (5) includes only outgoing (and evanescent) waves, i.e., here we use the Rayleigh hypothesis, [2], restricting the expansion to the terms with $z$-dependence of the form $\exp[-ip(q_n)z]$ only, and omitting those with $z$-dependence of the alternative form, $\exp[ip(q_n)z]$. This guarantees fulfillment of the boundary (radiation) conditions at $z = -\infty$.

The electric field possesses the $x$ and $z$ components only, $\mathbf{E} = (E_x, 0, E_z)$, and can be easily obtained from (3) and corresponding Maxwell equation. Specifically, the electric component of the diffracted field can be presented in the series of the form coinciding with that of $\mathbf{H}^D$,

$$\mathbf{E}^D = \sum_{n=-\infty}^{\infty} \mathbf{E}_n \exp[iq_n x - ip(q_n)z], \ z \le \zeta(x). \quad (5)$$

At the interface the total fields, $\mathbf{H} = \mathbf{H}^i + \mathbf{H}^D$, $\mathbf{E} = \mathbf{E}^i + \mathbf{E}^D$, are to satisfy the impedance boundary conditions, [14],

$$\mathbf{E}_t = \xi[\mathbf{n} \times \mathbf{H}] \text{ for } z = \zeta(x), \quad (6)$$

where the subindex $t$ denotes tangential to the interface component of the corresponding vector, $\xi$ denotes the surface impedance, and $\mathbf{n}$ stays for the unit vector normal to the interface directed into the dielectric, i.e., $\mathbf{n} = -[\mathbf{e}_z - \mathbf{e}_x \partial\zeta/\partial x]/\sqrt{1+(\partial\zeta/\partial x)^2}$.[3] We use Gauss units so that the surface impedance $\xi$ is dimensionless, and for nonmagnetic media $\xi = \sqrt{\varepsilon_d/\varepsilon}$.

The profile Fourier series expansion is

$$\zeta(x) = \sum_{n=-\infty}^{\infty} \zeta_n \exp(ingx), \quad (7)$$
$$g = 2\pi/d > 0, \ \zeta_{-n} = \zeta_n^*, \ \zeta_0 = 0.$$

The condition $\zeta_0 = 0$ corresponds to the specific choice of $Oz$ axis origin. The Fourier series coefficients of the interface normal, $\mathbf{n} = \mathbf{n}(x)$, can be expressed in terms of $\zeta_n$.

Substituting into Eq. (6) the fields representations given in Eqs. (4), (5), expressing the electric field Fourier amplitudes, $\mathbf{E}_n$, in terms of the magnetic ones, $H_n$, and equating terms with equal space dependence, we arrive at the infinite system of linear algebraic equations for the transformation coefficients (TCs), $h_n = H_n/H$,

$$\sum_{m=-\infty}^{\infty} D_{nm} h_m = V_n, \ n = 0, \pm 1, \pm 2, \ldots, \quad (8)$$

where the matrix of the system, $\hat{D} = \|D_{nm}\|$, and the right-hand side column vector, $\hat{V} = col\{V_n\}$, represent functionals depending on the problem parameters, specifically, the profile $\zeta(x)$. The coefficients of the system allow infinite series expansions with respect to $\zeta_n$. It is essential that strong diffraction anomalies take place for rather shallow gratings such that $k|\zeta|, |d\zeta/dx| \ll 1$, see [11-13] and below, so the expansions are very useful. For shallow gratings, we can restrict the series expansions of the coefficients to the main (linear) terms only, so that

$$D_{nm} = (\beta_n + \xi)\delta_{nm} - i(1 - \alpha_n \alpha_m)\mu_{n-m}, \ n,m = 0, \pm 1, \pm 2, \ldots, (9)$$
$$V_n = (\beta_n - \xi)\delta_{n0} + i(1 - \alpha_n \alpha_0)\mu_n, \ n = 0, \pm 1, \pm 2, \ldots. \quad (10)$$

Here $\delta_{nm}$ stays for the Kronecker delta-symbol, and

$$\mu_n = k\zeta_n, \ \alpha_n = \alpha + n\kappa, \ \kappa = g/k, \ \beta_n = \sqrt{1-\alpha_n^2},$$
$$\text{Re}, \text{Im } \beta_n \ge 0, \ n = 0, \pm 1, \pm 2, \ldots, \ n \in \mathbb{Z} \quad (11)$$

where $\alpha = \sin\theta$, $\theta$ denotes the incidence angle, $\mathbb{Z}$ stays for the set of integers.

Consider here the simplest (but of high interest) case of the grazing incidence, $0 < \beta \ll 1$ ($0 < 1 - \alpha \ll 1$). That is the specular reflected wave with necessity is the grazing one.

---

[3] The following considerations can be applied to the case of plane surface of metamaterials with periodically modulated electromagnetic properties such that the surface impedance is space-periodic, $\xi = \xi(x)$, $\xi(x+d) = \xi(x)$, cf. [11-13].

The simplest geometry of the problem is such, when only one of the diffracted waves except the specular wave is outgoing from the interface, all other diffraction orders correspond to evanescent waves. This geometry is presented in Fig. 1.

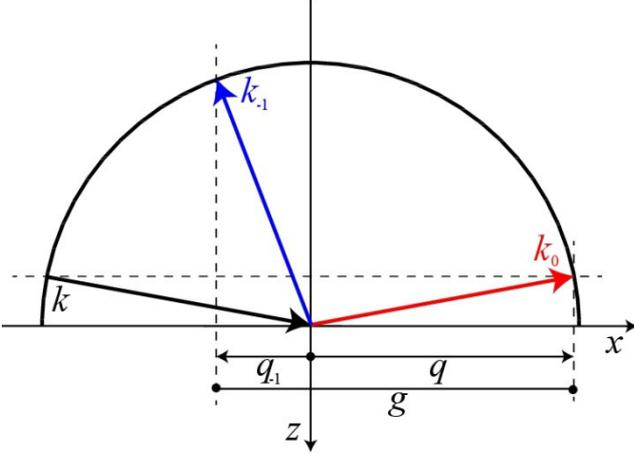

Fig. 1. Grazing incidence diffraction. The grating spacing, $d$, is supposed to be such that except the specular wave only the minus first diffraction order presents propagating wave, other diffraction orders correspond to evanescent ones, i.e., at $q \simeq k$, $q_1 = q + g > k$, $|q_{-1}| = |q - g| < k$, and $|q_n| > k$ for all $n \neq -1, 0$.

It should be emphasized, that among diffracted waves only the specular reflected one is close to the corresponding Rayleigh point, $\beta = \beta_0 \ll 1$, and all other waves are far enough from their branch points. That is, the only one diagonal element of the matrix $\hat{D} = \|D_{nm}\|$, namely, $D_{00} = \beta + \xi$, is small as compared with unity. Consequently, it is convenient to separate zeroth order equation and decompose the governing system, Eq. (8), as

$$\hat{\bar{D}}\hat{\bar{h}} = \hat{\bar{V}}, \tag{12}$$

$$D_{00} h_0 + \sum_{M \neq 0} D_{0M} h_M = V_0. \tag{13}$$

Here and below capital indexes denote all integers except zero,

$$\hat{\bar{D}} = \|D_{NM}\|, \quad N, M = \pm 1, \pm 2, \ldots, \tag{14}$$

$\hat{\bar{h}}$ and $\hat{\bar{V}}$ stay for the column vectors,

$$\hat{\bar{h}} = col\{h_M\}, \quad \hat{\bar{V}} = col\{\bar{V}_M\}, \quad M = \pm 1, \pm 2, \ldots, \tag{15}$$

$$\bar{V}_M = V_M - D_{M0} h_0, \quad M = \pm 1, \pm 2, \ldots. \tag{16}$$

Let us present Eq. (13) in a more explicit form as well,

$$(\beta + \xi) h_0 - i \sum_{M \neq 0} (1 - \alpha_0 \alpha_M) \mu_{-M} h_M = \beta - \xi. \tag{17}$$

The submatrix $\hat{\bar{D}}$ is diagonally dominat due to the fact that all nondiagonal elements are small as compared with unity and all diagonal ones are of order unity or greater. Thus, it can be easily inversed by means of the regular series expansion. Formally, we can express all nonspecular amplitudes, $h_M$, in terms of the given parameters of the system and unknown at this stage amplitude $h_0$ as follows,

$$\hat{\bar{h}} = \hat{\bar{D}}^{-1} \hat{\bar{V}}, \tag{18}$$

or, more explicitly,

$$h_M = \sum_{L \neq 0} \left[\hat{\bar{D}}^{-1}\right]_{ML} \bar{V}_L, \quad M = \pm 1, \pm 2, \ldots. \tag{19}$$

Taking into account that according to Eq. (17) and Eq. (10),

$$\bar{V}_L = V_L - D_{L0} h_0 = i(1 - \alpha_L \alpha_0) \mu_L + i(1 - \alpha_L \alpha_0) \mu_L h_0 = \\ = i(1 + h_0)(1 - \alpha_L \alpha_0) \mu_L, \quad L = \pm 1, \pm 2, \ldots \tag{20}$$

rearrange Eq. (19) as

$$h_M = i(1 + h_0) \sum_{L \neq 0} \left[\hat{\bar{D}}^{-1}\right]_{ML} (1 - \alpha_L \alpha_0) \mu_L, \quad M = \pm 1, \pm 2, \ldots \tag{21}$$

Substituting this expression into Eq. (17) we arrive at the closed linear equation for the specular TC,

$$(\beta + \xi) h_0 + (1 + h_0) \times \\ \times \sum_{M, L \neq 0} \left[\hat{\bar{D}}^{-1}\right]_{ML} (1 - \alpha_0 \alpha_M)(1 - \alpha_L \alpha_0) \mu_L \mu_{-M} = \beta - \xi. \tag{22}$$

Let, for brevity,

$$\Gamma = \sum_{M, L \neq 0} \left[\hat{\bar{D}}^{-1}\right]_{ML} (1 - \alpha_0 \alpha_M)(1 - \alpha_L \alpha_0) \mu_L \mu_{-M}, \tag{23}$$

$$\xi_{eff} = \xi + \Gamma. \tag{24}$$

Then the solution of Eq. (22) for $h_0$ can be presented as

$$h_0 = \frac{\beta - \xi_{eff}}{\beta + \xi_{eff}}. \tag{25}$$

It is of interest that the specular TC form, Eq. (25), coincides with the corresponding Fresnel coefficient $R$ related to the unmodulated (plane) interface,

$$R = \frac{\beta - \xi}{\beta + \xi}. \tag{26}$$

For the nonspecular TCs it follows identically,

$$h_M = \frac{2i\beta}{\beta + \xi_{eff}} U_M, \quad M = \pm 1, \pm 2, \ldots, \tag{27}$$

where the subsidiary functions $U_M$ are

$$U_M = \sum_{L \neq 0} \left[\hat{\bar{D}}^{-1}\right]_{ML} (1 - \alpha_L \alpha_0) \mu_L, \quad M = \pm 1, \pm 2, \ldots \tag{28}$$

It is essential that the coefficients $U_M$ experience only slow dependence on the parameters of interest in the vicinity of the point $\beta = 0$, as well as the functions $\Gamma$ and $\xi_{eff}$. Noteworthy, the quantity $\Gamma$ can be expressed in terms of $U_M$ as

$$\Gamma = \sum_M (1 - \alpha_0 \alpha_M) U_M \mu_{-M}. \tag{29}$$

In what follows we are dealing with rather smooth and shallow gratings so here we present the main terms of the necessary expansions. Let us introduce,

$$\hat{\bar{D}} = \hat{\bar{B}}\left(\hat{\bar{I}} - \hat{\bar{T}}\right), \quad (30)$$

where

$$\hat{\bar{I}} = \|\delta_{NM}\|, \quad \hat{\bar{T}} = \|\bar{T}_{NM}\|, \quad \hat{\bar{B}} = \|\bar{B}_{NM}\|, \quad (31)$$
$$N, M = \pm 1, \pm 2, \ldots$$

$$\bar{T}_{NM} = \frac{i}{b_N}(1 - \alpha_N \alpha_M)\mu_{N-M}, \quad \bar{B}_{NM} = b_N \delta_{NM},$$
$$b_N = \beta_N + \xi, \quad N, M = \pm 1, \pm 2, \ldots \quad (32)$$

Then

$$\hat{\bar{D}}^{-1} = \left(\hat{\bar{I}} - \hat{\bar{T}}\right)^{-1} \hat{\bar{B}}^{-1} = \left[\sum_{s=0}^{\infty} \hat{\bar{T}}^s\right] \hat{\bar{B}}^{-1}. \quad (33)$$

This series expansion converges under the condition $\left|\hat{\bar{T}}\right| < 1$, where $|\ |$ stays for the appropriate matrix norm. It is important that this condition is not restrictive: it allows consideration of strong anomalies, see [9, 11-13] and below. Moreover, strong anomalies hold for $\left|\hat{\bar{T}}\right| \ll 1$. So, it is sufficient to restrict our consideration to the first terms of the corresponding initial expansions. With accuracy up to the second-order terms with respect to $\mu$,

$$\left[\hat{\bar{D}}^{-1}\right]_{ML} = \left[\delta_{ML} + T_{ML} + \sum_{K \neq 0} T_{MK} T_{KL}\right] b_L^{-1} + O(\mu^3), \quad (34)$$

or, more explicitly,

$$\left[\hat{\bar{D}}^{-1}\right]_{ML} \simeq b_L^{-1}\left[\delta_{ML} + \frac{i}{b_M}(1 - \alpha_M \alpha_L)\mu_{M-L}\right] -$$
$$- \frac{b_L^{-1}}{b_M} \sum_{K \neq 0} \frac{1}{b_K}(1 - \alpha_M \alpha_K)(1 - \alpha_K \alpha_L)\mu_{M-K}\mu_{K-L}. \quad (35)$$

After simple rearrangement, we obtain the alternative expression,

$$\left[\hat{\bar{D}}^{-1}\right]_{ML} \simeq b_M^{-1}\left[\delta_{ML} + \frac{i}{b_L}(1 - \alpha_M \alpha_L)\mu_{M-L}\right] -$$
$$- b_M^{-1} \sum_{K \neq 0} \frac{1}{b_L b_K}(1 - \alpha_M \alpha_K)(1 - \alpha_K \alpha_L)\mu_{M-K}\mu_{K-L}. \quad (36)$$

Consequently, up to the second-order terms it follows from Eq. (28),

$$U_M \simeq b_M^{-1}(1 - \alpha_M \alpha_0)\mu_M +$$
$$+ i b_M^{-1} \sum_{L \neq 0} b_L^{-1}(1 - \alpha_L \alpha_0)(1 - \alpha_L \alpha_M)\mu_L \mu_{M-L}, \quad (37)$$
$$M = \pm 1, \pm 2, \ldots$$

Noteworthy, here the second-order terms are essential if the corresponding Fourier amplitude of the grating, $\mu_M$, vanishes or is anomalously small. Under this condition, the anomalous effects in $M$-th diffraction order are small and thus of low interest. Therefore, below we restrict our consideration to the linear term of $U_M$ expansion.

The main term of the quantity $\Gamma$ expansion is the square one,

$$\Gamma = \sum_{M \neq 0} b_M^{-1}(1 - \alpha_0 \alpha_M)^2 |\mu_M|^2. \quad (38)$$

Here we emphasize that the results obtained are valid for arbitrary angle of incidence for which all nonspecular reflected waves are far from their Rayleigh and resonance points, i.e., the inequality $|\beta_n| \gg |\xi|$ holds for the integers $n = \pm 1, \pm 2, \pm 3, \ldots$. Only the specular reflected wave can be arbitrary close to the grazing propagation. The subcase $\beta \ll 1$ is the specific one and its treatment cannot be carried out by means of the perturbation theory even for small grating height and inclinations, $|\mu| \ll 1$, $|\partial \zeta / \partial x| \ll 1$. The results presented above take into account this point. But they are evidently valid not only for $\beta \ll 1$, but for arbitrary angles of incidence, $0 \leq \beta \leq 1$, with single limitation indicated. Namely, the sufficient condition is $|\beta_n| \gg |\xi|$ for $n = \pm 1, \pm 2, \pm 3, \ldots$.

### 3. Energy flux extremes.

The solution obtained allows one to analyze in detail its dependence on the angle of incidence and all other parameters of the problem. Expressions (25), (29) describe the fast dependence of the TCs on the angle of incidence through the quantity $\beta = \cos\theta \ll 1$. Other functions entering the solution, $U_N$, $\xi_{eff}$, etc., are slow ones under the condition $\beta \ll 1$ (and far enough from other specific points $\beta_n \simeq 0$ for $n \neq 0$). Thus, for preliminary analytical considerations, these functions can be replaced by constants relating to their values at $\beta = 0$. This fact allows performing a thorough analytical investigation of the problem. Starting with the specular reflectivity, $\rho(\beta)$,

$$\rho(\beta) = |h_0|^2 = \frac{(\beta - \xi'_{eff})^2 + \xi''^2_{eff}}{(\beta + \xi'_{eff})^2 + \xi''^2_{eff}}, \quad (39)$$

one can see that it possesses specific minimal value at some point, $\beta = \beta_{extr}$, such that,

$$\beta_{extr} = |\xi_{eff}|. \quad (40)$$

In arriving to Eq. (40) we have neglected slow $\xi_{eff}$ dependence on $\beta$. With high accuracy, one can approximate $\xi_{eff}$ here and below by it value at $\beta = 0$. In Fig. 2 the specular reflectivity dependence on $\beta = \cos\theta$, ($\theta$ denotes the angle of incidence) is presented for harmonic gratings in the vicinity of the point $\beta = 0$.

At the extreme point, $\beta = \beta_{extr}$, $\rho$ exceeds its minimal value,

$$\rho = \rho_{\min}, \quad \rho_{\min} \equiv \rho(\beta_{extr}) = \frac{|\xi_{eff}| - \xi'_{eff}}{|\xi_{eff}| + \xi'_{eff}}. \quad (41)$$

Here and below the prime (double prime) denotes the real (imaginary) part of the corresponding quantity. The specular TC field at the point $\beta = \beta_{extr}$ is as follows,

$$h_0(\beta_{extr}) = \frac{|\xi_{eff}| - \xi_{eff}}{|\xi_{eff}| + \xi_{eff}}. \quad (42)$$

The $\rho$ dependence on the angle of incidence in terms of the variable $\beta = \cos\theta$ is illustrated in Fig. 2. As it strictly follows from Eqs. (40), (38), (24), and it is easy to see from Fig. 2, the position of the $\rho$ minimum shifts toward greater $\beta$ values and the minimum widens and deepens with the grating depth increase.

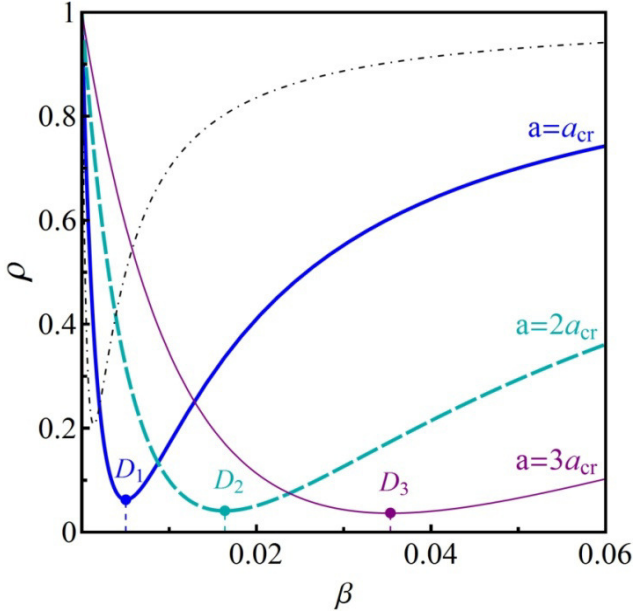

Fig. 2. The specular intensity $\rho$ dependence on the $\beta = \cos\theta$ is presented for three Cu harmonic gratings of equal period $d = 200$ mkm and different depths $a$ indicated near the curves, and for the plane interface (dash-and-dot curve). The calculations were performed for the wavelength $\lambda = 300$ mkm ($\xi = 0.0009 - 0.001i$, [19]) so that the characteristic grating dimensionless parameters are $\kappa = 1.5$ and $a_{cr} = 0.037$. The points $D_n$, correspond to the $\rho$ minimal values at $a = n \cdot a_{cr}$, $n = 1, 2, 3$, and are as follows: $D_1 = (0.005, 0.063)$, $D_2 = (0.016, 0.041)$, $D_3 = (0.035, 0.037)$.

This is also clearly illustrated in Fig. 3, where the color grade demonstrates the value of specular intensity $\rho$ for grating depths from $a = a_{cr}$ to $a = 5a_{cr}$ and $\beta$ values from 0 to 0.06. One can notice, that with the increase of the grating depth the value of $\beta$ where the minimum is observed shifts towards greater $\beta$. This shift is demonstrated by points $D_1$, $D_2$ and $D_3$ the dotted line in the plot. Also, the region that corresponds to the lower values of $\rho$ (darker region) enlarges with the increase of the grating depth. Thereby, comparing Fig. 3 and Fig.2 one can conclude that for shallow gratings we received sharp minimum, while for the deep gratings we obtain wide minimum.

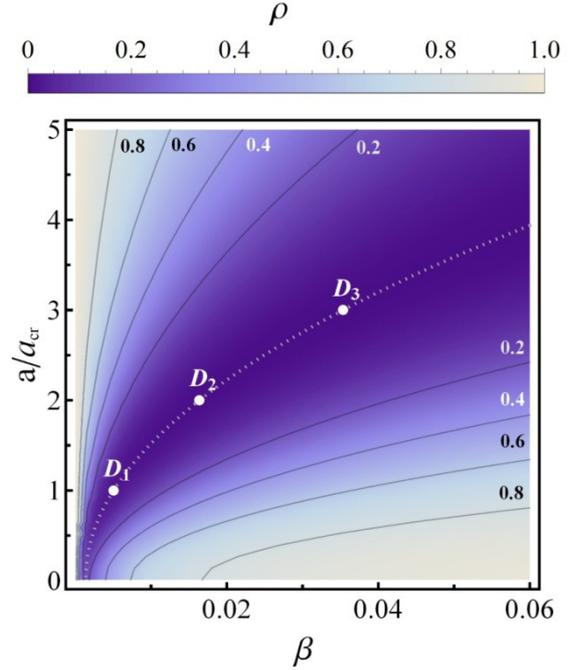

Fig. 3. The $\rho$ dependence on the grating depth and $\beta$ for Cu grating $\lambda = 300$ mkm, $\xi = (0.0009 - 0.001 i)$, $a_{cr} = 0.037$. The dotted line corresponds to $\rho$ minimal values; $\rho$ contour curves are shown by solid lines. The points $D_n$, $n = 1, 2, 3$, are the same as in Fig. 2.

Noteworthy, the reflectivity minimum is of rather general character and exists even for TM polarized wave incidence on unmodulated interfaces, $\Gamma = 0$, $\xi_{eff} \Rightarrow \xi$ (when $h_0$ coincides with the corresponding Fresnel reflection coefficient $R$). The corresponding discussion can be found in the textbook [14]. This minimum along with that under discussion is analogous to the reflectivity minimum from dielectric media existing under Brewster angle incidence (when the reflected and transmitted waves are propagating at a right angle), [14]. In view of the fact that for $|\varepsilon| \gg 1$ (which is typical for good metals up to the frequencies of the visible range), the normal to the interface component of the wavevector in the metal half-space prevails essentially the tangential one, so the wave in the metal region can be formally considered as orthogonal to the interface. Consequently, under grazing incidence the reflected from the metal wave is approximately orthogonal to the "transmitted" one. In other words, the suppression of the specular reflectivity for TM polarization at grazing incidence is analogous to that at the Brewster angle. Recall, the Brewster angle of incidence from the vacuum, $\theta_{Br}$, is defined for dielectric media as

$$\sin\theta_{Br} = \sqrt{\varepsilon}/\sqrt{\varepsilon+1},$$

so that for metals when $\varepsilon$ is complex-valued, but $|\varepsilon| \gg 1$, one formally obtains from this relation $\theta_{Br} \simeq \pi/2$.

The specular reflectivity minimum, Eq. (41), becomes deep for relatively high effective losses, i.e., for $\xi'_{eff}$ comparable to $|\xi_{eff}|$ (see Fig. 2). It worth pointing out here that $\xi'_{eff}$ includes both the dissipative and radiative losses relating to the quantities $\xi'$ and $\Gamma'$, respectively. The quantity $\Gamma'$ is mainly caused by the outgoing (propagating) waves.

On the contrary, $\rho_{min}$ approaches unity at vanishing losses, $\xi'_{eff} \to 0$. Therefore, the effect of the specular reflection suppression under consideration is mainly attributed to the cumulative (both active and radiative) losses maximum, cf. [15, 16]. However, as it is shown below, the point $\beta = \beta_{extr}$ corresponds not only to the specular reflection minimum but results in well expressed maximal nonspecular efficiencies along with the active losses maximum. Evidently, if the only propagating diffracted wave is the specular one, then the grazing minimum is with necessity accompanied by maximal absorption. Noteworthy, the reflectance minimum under grazing incidence can correspond to the essential redirection of the energy into the nonspecular diffraction channels corresponding to propagating waves even for shallow gratings. Below this thesis is illustrated for the simplest case when in addition to the specular wave only one diffracted order corresponds to the propagating (outgoing) wave. It can be realized if $1+\alpha > \kappa > 1$, when the minus first order presents propagating wave, $\beta_{-1} > 0$, and $\beta_n$ with $n \neq -1, 0$ are pure imaginary. In what follows we consider that $\beta_{-1}$ is of order unity, so that the minus first diffraction order is far from its Rayleigh point.[4]

---

[4] Alternative case resulting in strong GA is of interest too, as well as the specific case when GA is accompanied by SPP anomaly relating to some other diffraction order. These cases correspond to the double and combined anomalies and will be considered in forthcoming papers.

It is of interest that normalized intensities of the propagating diffraction orders,

$$\rho_N = |h_N|^2 \frac{\mathrm{Re}(\beta_N)}{\beta} = 4W|U_N|^2 \mathrm{Re}(\beta_N), \quad (43)$$

present strongly nonmonotonic $\beta$ functions in accordance with the fast dependence of the subsidiary function introduced, $W = W(\beta)$,

$$W(\beta) = \frac{\beta}{(\beta + \xi'_{eff})^2 + (\xi''_{eff})^2}. \quad (44)$$

It is easy to see that $W(\beta)$ achieves its maximal value, $W_{max}$, strictly at the point $\beta = \beta_{extr}$ and is high,

$$W_{max} = W(\beta_{extr}) = \frac{1}{2(|\xi_{eff}| + \xi'_{eff})} \gg 1. \quad (45)$$

That is when all intensities of all propagating waves (except the specular one) simultaneously achieve their maximal values at the point $\beta = \beta_{extr}$,

$$\rho_{N,max} = \rho_N(\beta_{extr}) = \frac{2|U_N|^2}{|\xi_{eff}| + \xi'_{eff}} \mathrm{Re}(\beta_N), \quad (46)$$
$$N = \pm 1, \pm 2, \ldots$$

This property is illustrated in Fig. 4, where the incident angle dependence of the minus first diffraction order intensity, $\rho_{-1}$, is shown for the geometry of Fig. 1.

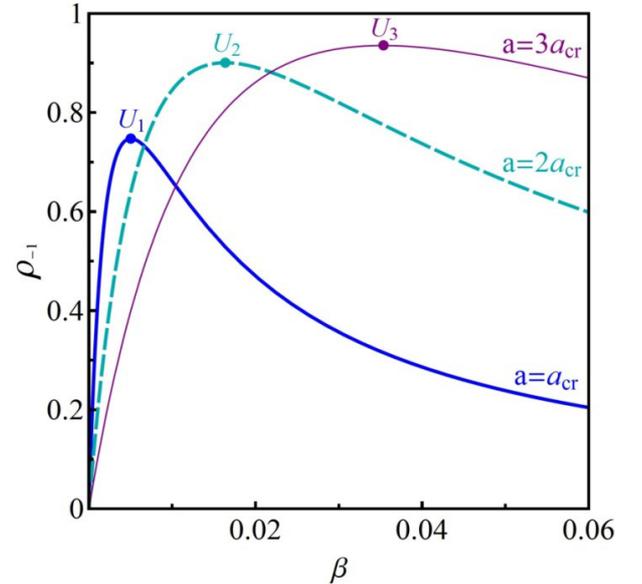

Fig. 4. The $\rho_{-1}$ plot versus $\beta$ for three harmonic gratings differing by depth with parameters indicated in Fig. 2. The points $U_n$ correspond to the maxima related to gratings with $a = n \cdot a_{cr}$, $n = 1, 2, 3$, respectively, and are as follows: $U_1 = (0.005, 0.749)$, $U_2 = (0.016, 0.902)$, $U_3 = (0.035, 0.937)$.

In Fig. 5 the $\rho_{-1}$ dependence on $\beta$ and the grating height is demonstrated for the conditions coinciding with those of Fig. 3. It illustrates not only the anomaly shift to greater $\beta$ and widening with the grating height increase, but also the $\rho_{-1}$ value in a wide range of the crucial parameters.

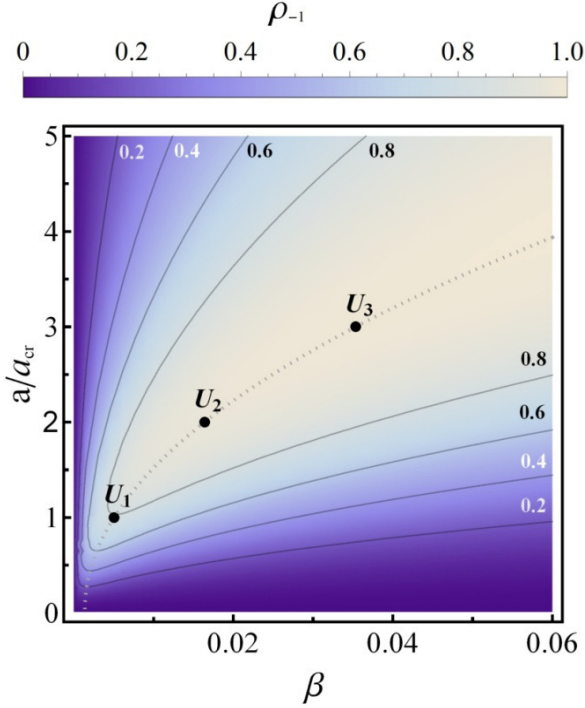

Fig. 5. The $\rho_{-1}$ dependence on the grating depth and $\beta$ for Cu gratings at $\lambda=300$ mkm, $\xi = (0.0009 - 0.001\,i)$, $a_{cr} = 0.037$. The dotted line corresponds to maximal values, solid lines depict $\rho_{-1}$ contour curves with levels indicated.

One can see from Fig. 6 as well as from Fig. 2 and Fig. 4 that the positions of the $\rho_{-1}$ maxima coincide with high accuracy with those of $\rho$ minima. In addition, this point corresponds to the maximum of the absorption $A$ discussed in detail below.

The total energy flux outgoing with the propagating waves does not exceed that of the incident wave, i.e.,

$$\sum_N \rho_N \le 1,$$

where $\rho_0$ stays for $\rho$. The difference between the sum and unity, $A = 1 - \sum_N \rho_N$, is nothing else than the active losses per unit area. The inequality for the solution presented is to be true under rather general conditions, specifically for such $\beta$ and $\kappa$ values that are far from anomalies related to all diffraction orders except the specular one. If the active losses are absent, then the inequality transforms into the equality. In the specific case of short-period gratings, such that $\kappa > 2$, all diffracted orders except zeroth one with necessity correspond to evanescent waves. Under such conditions the strong specular reflectivity suppression is accompanied by maximal absorption. The energy redistribution between outgoing waves and the dissipation strongly depends on the parameters of the problem, as one can see from the explicit solution. Specifically, in the geometry shown in Fig. 1 the absorption, $A = 1 - (\rho + \rho_{-1})$ for harmonic grating can be presented explicitly as

$$A = \frac{4\xi'\beta}{\left(\beta + \xi'_{eff}\right)^2 + \xi''^{2}_{eff}}.$$

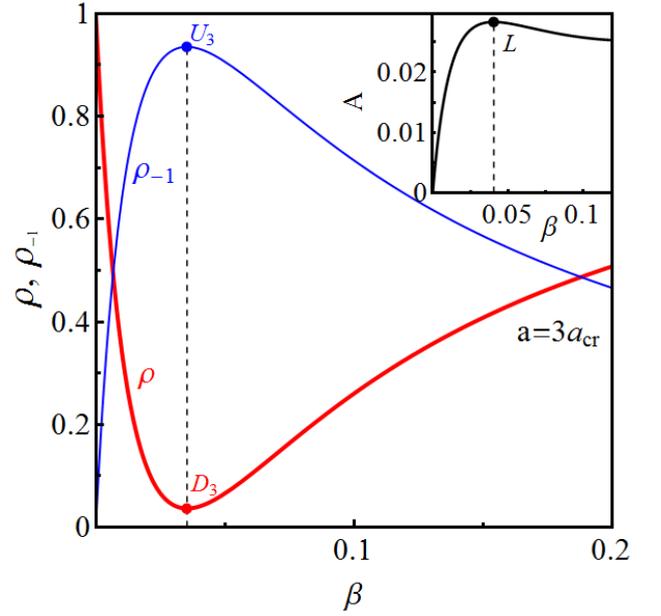

Fig. 6. The dependencies of the intensities $\rho$, $\rho_{-1}$, and of the absorption, $A = 1-(\rho + \rho_{-1})$, on $\beta$ for Cu grating at $\lambda=300$ mkm, $\xi = (0.0009 - 0.001\,i)$, and the grating depth $a = 3a_{cr}$, $a_{cr} = 0.037$, $\kappa = 1.5$. The values of $D_3$ and $U_3$ are the same as indicated in Fig. 2 and Fig. 4. The absorption maximum, $L$, is very low, $L = (0.04, 0.02)$, approximately all energy is redistributed to the minus first diffraction order.

It can be easily checked that $A$ possesses single maximum. Neglecting slow $\xi_{eff}$ dependence on $\beta$ one can make sure that the maximum is at the point $\beta = \beta_{extr}$, and is

$$A_{max} = \frac{2\xi'}{|\xi_{eff}| + \xi'_{eff}},$$

cf. point $L$ in Fig. 6. Evidently, the absorption vanishes if the medium is dissipation free, $\xi' = 0$. Under rather specific conditions $A_{max}$ can be of order unity, that does not describe general case contrary to the statement in [16].

For the case shown in Fig. 1 only two diffraction orders correspond to the propagating waves – the specular and the minus first ones. Consider this specific subcase in more detail. Suppose additionally that the grating is harmonic one, i.e.,

$$\mu(x) = 2a\cos(gx); \quad \mu_1 = \mu_{-1} = a > 0, \\ \mu_n = 0 \text{ for } |n| \geq 2 \quad (47)$$

Then, approximating $b_{\pm 1}$ by $\beta_{\pm 1}$, and taking into account that $\beta_{\pm 1}^2 \simeq -\kappa^2 \mp 2\kappa$ at the point $\beta = 0$, we find

$$\Gamma \simeq \kappa^2 a^2 \left[ \frac{1}{\sqrt{\kappa(2-\kappa)}} - i\frac{1}{\sqrt{\kappa(2+\kappa)}} \right]. \quad (48)$$

In the view of the fact that the specular reflectivity possesses rather expressed minimum, then for relatively low active losses the incoming energy is to be redirected into other propagating waves. The most interesting case that allows obtaining rather strong grazing anomalies presents such one that,

$$|\Gamma| \gg |\xi|, \quad (49)$$

but $|\Gamma| \ll 1$, i.e., the case when the effective impedance is mostly caused by the diffraction rather than by the medium properties[5]. It is of the essence that the supposition presented in Eq. (49) does not contradict the shallow character of the grating, $|\Gamma| \sim a^2 \ll 1$, in view of the surface impedance smallness, $|\xi| \ll 1$. The characteristic value of the dimensionless grating height, $a_{cr}$, defined so that for $a = a_{cr}$ $|\Gamma| \sim |\xi|$, is small, $a_{cr} = \sqrt{|\xi|} \ll 1$. Under Eq. (49) condition (equivalent to $1 \gg a \gg a_{cr}$) $\rho_{-1,\max}$ and $\rho_{\min}$ can be rewritten as

$$a \gg a_{cr}: \quad \rho_{-1,\max} \simeq \frac{2\Gamma'}{\beta_{-1}^2(|\Gamma|+\Gamma')}, \quad \rho_{\min} \simeq \frac{|\Gamma|-\Gamma'}{|\Gamma|+\Gamma'}, \quad (50)$$

or, in view of Eq. (48),

$$a \gg a_{cr}: \quad \rho_{-1,\max} \simeq \frac{2\sqrt{2+\kappa}}{2+\sqrt{2+\kappa}} \equiv \rho_{-1,\lim}, \\ \rho_{\min} \simeq \frac{2-\sqrt{2+\kappa}}{2+\sqrt{2+\kappa}} \equiv \rho_{-1,\lim}, \quad \rho_{-1,\lim} + \rho_{\lim} \simeq 1 \quad (51)$$

That is, for the rather deep gratings, such that $a \gg a_{cr}$ (but still $a \ll 1$) the energy redistribution does not depend on the grating height, the quantities $\rho_{-1,\max}$ and $\rho_{\min}$ achieve their asymptotic values depending on the geometrical parameters and the wavelength through the dimensionless combination $\kappa = \lambda/d$ only, Eq. (51).

It is easy to see that within the accuracy indicated $\rho_{-1,\max} + \rho_{\min} = 1$. That is, all incident energy is redistributed only between two propagating waves, when the active losses are negligible as compared with the radiation ones. This property takes place for rather deep gratings, such that $a_{cr} \ll a \ll 1$. Note also that for our conditions in general case $0 < 1 - (\rho_{-1} + \rho) \ll 1$, where the difference is caused by the active losses and vanishes with it vanishing.

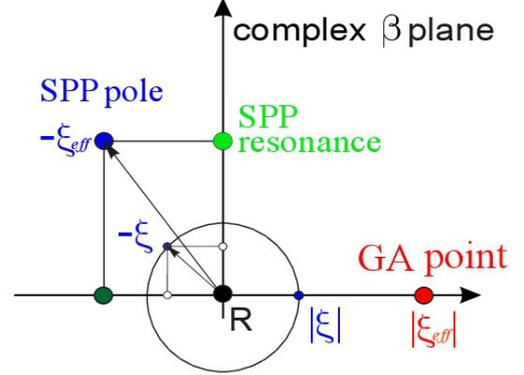

Fig. 7. Beta plane for some diffraction order. Only vicinity of the corresponding Rayleigh point (that is of main interest in view of the diffraction anomalies), $|\beta| \ll 1$, is shown. With the change of the parameters of the problem, the $\beta$ value for each diffraction order can be either pure real positive (propagating wave) or pure imaginary (evanescent wave). The exception here presents the case when $\beta$ corresponds to the incident wave, so that $\beta$ is pure real, $0 < \beta \leq 1$. These cases are separated by the Rayleigh (branch) point, R, $\beta = 0$. Other characteristic points, $\beta_{SPP} = -i\,\text{Im}(\xi_{\text{eff}})$ and $\beta_{GA} = |\xi_{\text{eff}}|$ related to the SPP resonance and to the grazing anomaly (GA), respectively, are shown by circles. If $\beta$ corresponds to the incident wave, then it is pure real and only GA point is actually of interest.

To the best of our knowledge, the grazing diffraction anomaly under discussion was not considered earlier. However, in [6, 15] one can find the related anomalous effect arising for such parameters of the diffraction problem that some diffracted order corresponds to the grazing wave propagating at the specific grazing angle. The anomaly consists of highly enhanced efficiency of this wave accompanied by deep suppression of the specular reflection. It is worth noticing that this grazing wave enhancement (GWE) is related to the problem under consideration by the reciprocity theorem, [17, 18]. Namely, reversing the propagation direction of the minus first order diffracted wave in Fig. 1 we arrive at the reciprocal diffraction problem. In the latter the corresponding minus first order is related to the grazing wave propagating in the opposite direction to the incident wave in the primordial problem. In more detail the reciprocity approach will be discussed in forthcoming papers.

---

[1] [5] Specifically, such condition holds within approximation of ideal metal, $\xi \to 0$, that is valid in the long wavelength region.

We also illustrate locations of other anomalous diffraction points related to the interface of metal and isotropic dielectric (vacuum, for simplicity). It is convenient to consider the effects in terms of the dimensionless normal component $\beta$ of the corresponding diffraction order. This quantity can be pure real or pure imaginary belonging to positive half-axis for both cases. The point $\beta = -\xi_{eff}$ in the $\beta$ plane, Fig. 2, shows corresponding diffraction order pole caused by the surface plasmon polariton (SPP) mode. Note, the specific value of $\xi_{eff}$ for a given grating depends on the "resonance" diffraction order, see Eqs. (23), (24).

## 4. Conclusion.

It is shown that the diffraction of TM polarized wave at the high reflecting gratings under grazing incidence can result in deep suppression of the specular reflection accompanied by considerable redirection of the incoming energy to other propagating diffracted waves. The detailed theoretical analysis of the problem is presented basing on the appropriate analytical approach. The diffraction anomaly considered in the paper is of general character and can take place for other wave types under appropriate conditions (high contrast of the adjacent media properties). In particular, it can be present at the interface of ordinary dielectric media and for left-handed media as well. The analogous anomaly does exist and is well expressed for magnetic high-contrast media interface for TE polarization.

### Acknowledgements

The authors wish to acknowledge the constructive comments of I. S. Spevak.